\documentclass[prl,twocolumn,superscriptaddress,showpacs,amsmath,amssymb]{revtex4-1}

\usepackage{graphicx}
\usepackage{latexsym}
\usepackage{amsmath}
\usepackage{amssymb}
\usepackage{amsfonts}
\usepackage{color}
\usepackage{bm}
\usepackage{verbatim}

\definecolor{MS-color}{RGB}{128,0,128}

\usepackage{framed}
\definecolor{shadecolor}{RGB}{222,222,221}
\begin{document}

\title{Thermally induced spin-transfer torques in superconductor/ferromagnet bilayers}

\author{I. V. Bobkova}
\affiliation{Institute of Solid State Physics, Chernogolovka, Moscow reg., 142432 Russia}
\affiliation{Moscow Institute of Physics and Technology, Dolgoprudny, 141700 Russia}
\affiliation{National Research University Higher School of Economics, Moscow, 101000 Russia}

\author{A. M. Bobkov}
\affiliation{Institute of Solid State Physics, Chernogolovka, Moscow reg., 142432 Russia}

\author{Wolfgang Belzig}
\affiliation{Fachbereich Physik, Universit{\"a}t Konstanz, D-78457 Konstanz, Germany}

\date{\today}


\begin{abstract}

Thermally induced magnetization dynamics is currently a flourishing field of research due to its potential application in information technology. We study the paradigmatic system of a magnetic domain wall in a thermal gradient which is interacting with an adjacent superconductor.  The spin-transfer torques arising in this system due to the combined action of the giant thermoelectric effect and the creation of equal-spin pairs in the superconductor are large enough to give rise to high domain wall velocities $10^3$ times larger than previously predicted.

\end{abstract}

\maketitle

In recent years a new  field of research has emerged by coupling spin and heat degrees of freedom -- called spin caloritronics \cite{Bauer2012}. In particular, the spin Seebeck effect, that is the generation of a spin imbalance by a temperature gradient, has been discussed. Further, a thermally induced spin-transfer torque (STT), based on the spin-dependent Seebeck effect was predicted and its influence on the domain wall (DW) motion has been  discussed\cite{Berger1985,Jen1986,Hatami2007,Kovalev2009,Hals2010,Hinzke2011,Yan2011,Moretti2017}. There is also experimental evidence of the thermally-induced STT in ferromagnetic systems via observations of the magnetization switching and domain wall motion  \cite{Jen1986_2,Torrejon2012,Jiang2013,Ramsay2015,Yu2010,Pushp2015}. The main mechanisms of the thermal STT are spin transfer via magnons and via thermally-induced electron spin flow. 

\begin{figure}[t]
    \centerline{\includegraphics[clip=true,width=0.8\columnwidth]{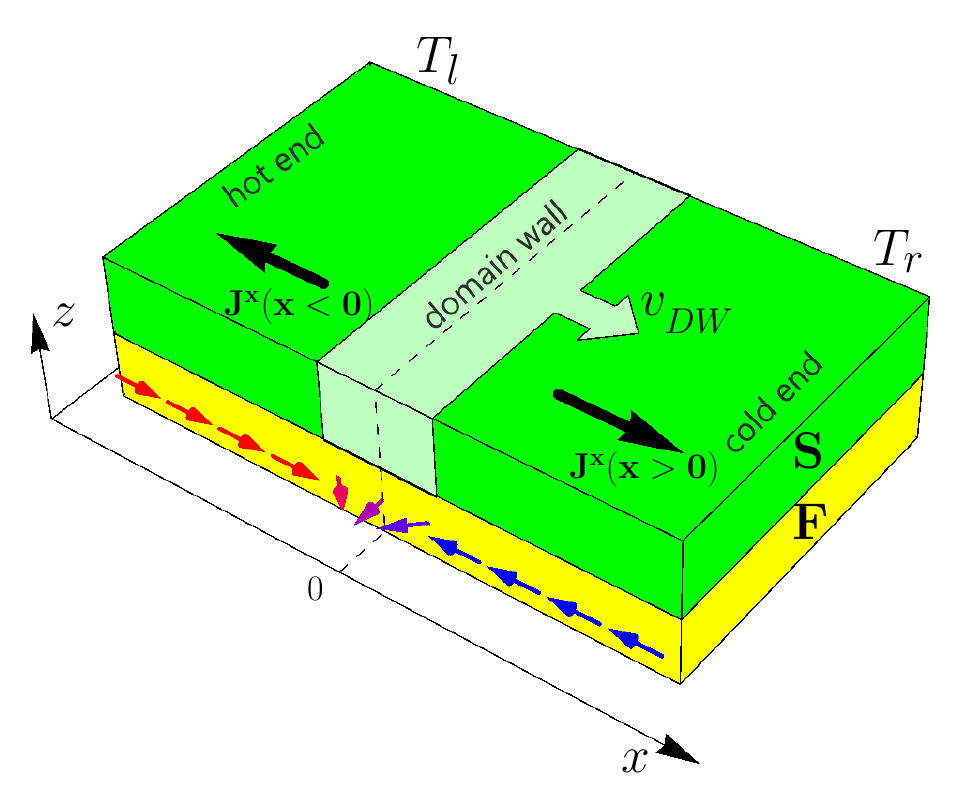}}
    \caption{Sketch of the bilayer S/F system. The magnetization of the ferromagnet F  has  a  form  of  a  head-to-head  domain  wall  (DW) and is indicated by arrows.  The picture on the top surface illustrates the process of thermally induced spin pumping into the DW region. Thermally induced quasiparticles both electron- and hole-like move from the hot to the cold end.  In the bulk of both domains the magnetic moments of the quasiparticles are polarized along the corresponding magnetization. Therefore, the spin current (opposite to the magnetization current) in the bulk of both domains is directed away from the DW. In the left (hotter) domain the direction of the majority spin flow is opposite to the spin current direction, while in the right (colder) domain they coincide. Therefore, the spin current flowing in both domains, pumps majority spins of the hotter domain into the DW region. This leads to the expansion of the hotter domain and, consequently, the DW moves from the hot to the cold end.}
 \label{sketch}
 \end{figure}

In this Letter, we propose a  paradigm of converting thermal gradients to magnetization dynamics in a very efficient and energy saving way. The key idea is to exploit a superconductor/ferromagnet hybrids as shown in Fig.~\ref{sketch}, where the STT is due to the combined action of the giant thermoelectric effect and creation of equal-spin pairs in the superconductor. Our estimates suggest that domain wall velocities of the order of at least a hundred m/s can be achieved by extremely small temperature differences smaller than the critical temperature of conventional low-temperature superconductors like Al and Nb. The efficiency of the thermal STT can be quantified by the ratio of the domain wall velocity $v_{DW}$ to the temperature gradient $\nabla T$. In principle, our estimates give $v_{DW}/\nabla T \gtrsim 10-10^2$ mm$^2$/Ks for the S/F system, which is about three order of magnitude larger than the values $\sim 10^{-2}-10^{-1}$ mm$^2$/Ks reported for thermally induced domain wall motion in ferromagnetic materials\cite{Jen1986_2,Jiang2013}.  

In the framework of the discussed mechanism the thermal STT is provided by the electron spin polarization created in {\it the superconducting part} of the structure and subsequent coupling of this polarization to the ferromagnet magnetization via the exchange mechanism. In principle, the STT is universal and can be relevant for ferromagnetic metals as well as for magnetic insulators. The key ingredient for efficient realization of the thermal STT is the Zeeman splitting of the DOS in the superconductor by proximity to the adjacent ferromagnet. The role of the Zeeman splitting is twofold.
First,  the  presence  of superconductivity in the system provides a unique mechanism for an anti-damping spin-transfer torque, which is not connected to spin-flip scattering of quasiparticles: a superconducting quasiparticle spin cannot align itself to the inhomogeneous magnetization at the length scales shorter than the superconducting coherence length $\xi_S$ because  the quasiparticle strongly interacts with the condensate of equal-spin pairs, where the characteristic length scale is $\xi_S$. Therefore, if the DW width $l_{DW}$ is less than $\xi_S$, the quasiparticle spins are inevitably misaligned to the DW magnetization giving rise to the non-adiabatic torque.

The second contribution to the STT is the thermally-induced quasiparticle spin flow in the superconducting part (spin-dependent Seebeck effect). This quasiparticle spin flow is known to result in the thermal STT in nonsuperconducting systems \cite{Berger1985,Jen1986,Hatami2007,Kovalev2009,Hals2010,Hinzke2011,Yan2011,Moretti2017}.  
At the same time a Zeeman-split superconductor is a unique example of a physical system, in which a very large spin Seebeck effect can be realized at low temperatures. The spin Seebeck effect can be quantified in terms of the spin thermopower $\nabla \mu_s/2e\nabla T$ generated by the temperature gradient $\nabla T$ in an open circuit, where $\mu_s = \mu_\uparrow - \mu_\downarrow$ is the spin imbalance of the spin-dependent chemical potentials.  In order to have a nonzero spin Seebeck effect an electron-hole asymmetry at the Fermi level is required \cite{Johnson87,Gravier2006,Uchida2008,Uchida2010,Bauer2012,Slachter2010}. Typically the corresponding electron-hole asymmetry in metallic ferromagnets is rather small resulting in the predicted spin thermopower $\mu_s/2e\nabla T \sim 10^{-3}$mV/K for intermetallic interfaces at room temperature \cite{Hatami2007,Hatami2009}, while a much smaller spin thermopower $\sim 10^{-6}$mV/K was measured for a ferromagnetic film \cite{Uchida2008}. In Zeeman-split superconductors a very large thermopower and spin thermopower was predicted  \cite{Machon2013,Kalenkov2012,Ozaeta2014,Giazotto2014,Kalenkov2015,Machon2014,Kawabata2013,Giazotto2015,Linder2016,Bobkova2017,Rezaei2018,Aikebaier2018_2}.  The experimental observation of the large thermopower has been reported \cite{Kolenda2016,Kolenda2017,Kolenda2016_2}. Upon application of strong in-plane magnetic
fields $B \sim 1$T or by proximity to a magnetic insulator, Seebeck coefficients of the order of $0.3$mV/K
were measured, which is comparable to the thermopower measured in magnetic semiconductors at much higher temperatures\cite{Pu2008}. 

{\it Model and method.} The model system that we consider is shown in Fig.~\ref{sketch}. It consists of a spin-textured ferromagnet with a spatially dependent magnetization $\bm M(\bm r)$ in contact to a spin-singlet superconductor. The superconductor is assumed to be in the ballistic limit. The ferromagnet can be a metal or an insulator.  If the thickness of the S film $d_S$ is smaller than the superconducting coherence length $\xi_S$, the magnetic proximity effect, that is the influence of the adjacent ferromagnet on the S film can be described by adding the effective exchange field\cite{Bulaevskii1982,Tokuyasu1988,Millis1988,Belzig2000,Bergeret2005,Buzdin2005} $\bm h(\bm r) \sim -\bm M(\bm r)$ to the quasiclassical Eilenberger equation, which we use below to treat the superconductor. While in general the magnetic proximity effect is not  reduced to the effective exchange only \cite{Cottet2009,Eschrig2015_2,Kamra2018}, in the framework of the present study we neglect other terms which can be viewed as additional magnetic impurities in the superconductor and focus on the effect of the spin texture.
The bilayer film is assumed to be connected to  equilibrium reservoirs having different temperatures $T_{l,r}$. We neglect all inelastic relaxation processes in the film assuming that its length is shorter than the corresponding relaxation length. 

The torque can be calculated starting from the effective exchange interaction between the spin densities on the two sides of the S/F interface:
\begin{eqnarray}
H_{int} = - \int d^2 \bm r J_{ex} \bm S \bm s,
\label{interface_ham}
\end{eqnarray}
where $\bm s$ is the electronic spin density operator in the S film, $\bm S$ is the localized spin operator in the F film, $J_{ex}$ is the exchange constant and the integration is performed over the 2D interface. It has been shown \cite{Kamra2018} that this exchange interaction Hamiltonian results in the appearance of the exchange field $h = J_{ex} M/(2\gamma d_s)$ in the S film. Here $M$ is the saturation magnetization of the ferromagnet and $\gamma$ is the gyromagnetic ratio.

The spin density $\bm s$ obeys the following equation:
\begin{eqnarray}
\partial_t \bm s = - \partial_j \bm J_j - 2 \bm h \times \bm s,
\label{electron_spin}
\end{eqnarray}
where we have introduced the vector $\bm J_j = (J_j^x,J_j^y,J_j^z)$ corresponding to the spin current flowing along the $j$-axis in real space.

The additional contribution to the Landau-Lifshitz-Gilbert equation from the exchange interaction Eq.~(\ref{interface_ham}) has the form of a torque acting on the magnetization:
\begin{eqnarray}
\frac{\partial\bm M}{\partial t} = -\gamma \bm M \times \bm H_\textrm{eff} + \frac{\alpha}{M} \bm M \times \frac{\partial\bm M}{\partial t} + \frac{J_{ex}}{d_F} \bm M \times \bm s,~~~~~~
\label{LLG}
\end{eqnarray}
where $\alpha$ is the Gilbert damping constant and the last term represents the torque. $\bm H_\textrm{eff}$ is the local effective field
\begin{eqnarray}
\bm H_\textrm{eff} = \frac{H_K M_x}{M}\bm e_x + \frac{2A}{M^2}\nabla^2 \bm M - K_\perp M_z \bm e_z  .~~~~~~
\label{H_eff}
\end{eqnarray}
$H_K$ is the anisotropy field, along the $x$-axis, $A$ is the exchange constant and the self-demagnetization field $K_\perp M_z$ is included.

In a stationary situation $\partial_t \bm s = 0$ from Eq.~(\ref{electron_spin}) one can obtain that
\begin{eqnarray}
\bm N = \frac{J_{ex}}{d_F} \bm M \times \bm s = \gamma \frac{d_S}{d_F} \partial_j \bm J_j. 
\label{torque}
\end{eqnarray}
The spin current $\bm J_j$ in the superconductor is calculated in the framework of the Keldysh technique for quasiclassical Green's functions. All the technical details of the Green's function calculation are given in the Supplementary Material.

To understand the efficiency of the torque $\bm N$ induced by the presence of the superconductor, we compare its value to the characteristic value of the torque induced by the effective field $H_\textrm{eff}$. Eq.~(\ref{torque}) can be rewritten as $\bm N/\gamma H_K M = \zeta \partial_{\tilde x} \tilde {\bm J}_x$, with the dimensionless quantities $\partial_{\tilde x} \tilde {\bm J}_x = (2e^2 R_N v_F/\Delta_0^2) \partial_x \bm J_x$ and $\zeta = E_S/\pi E_A$. The latter is proportional to the ratio of the condensation energy $E_S = N_F \Delta_0^2 d_S/2$ and the anisotropy energy $E_A = M H_K d_F/2$ per unit area of the film in the $(x,y)$-plane. Here and below $R_N = \pi/(2e^2 N_F v_F)$ is the normal state resistance of the film and $\Delta_0$ is the superconducting order parameter of the S film in the absence of the ferromagnet at zero temperature. Taking $E_S \sim  d_S \times (10 \div 10^3)\;\textrm{erg/cm}^3$ (for conventional superconductors like Al and Nb) and $E_A \sim d_F \times 10^5\;\textrm{erg/cm}^3$ for Py thin films \cite{Beach2005,Beach2006} or $E_A \sim d_F \times (10 \div 10^2)\;\textrm{erg/cm}^3$ for YIG thin films \cite{Mendil2019}, we obtain that $\zeta$ can vary in a wide range $\zeta \sim (10^{-4} \div 10^2)(d_S/d_F)$.

\begin{figure}[t]
    \centerline{\includegraphics[clip=true,width=3.2in]{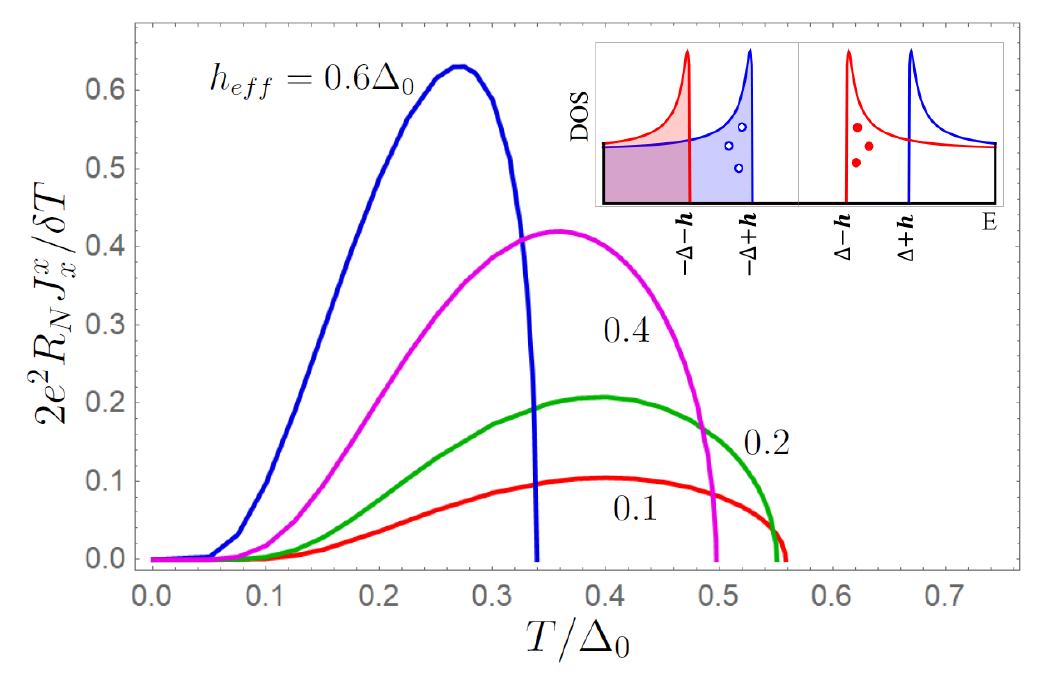}}
    \caption{Spin current divided by the temperature difference $\delta T \to 0$ in the homogeneous S/F bilayer vs the temperature. $h_{eff}=0.1$(red), $0.2$(green), $0.4$(purple), $0.6$(blue) in units of $\Delta_0$. Insert: spin resolved DOS filled by thermally activated right-moving quasiparticles coming from the hot end. The spin-up $S_x=+1$ (spin down $S_x=-1$) DOS is blue (red). It is seen that all the right-moving quasiparticles contribute to spin flow of the same direction.}
 \label{bulkspin}
 \end{figure}

{\it Thermally induced spin current in a homogeneous S/F bilayer.} Now we are ready to calculate the thermally induced spin torque in the S/F bilayer. But at first we discuss briefly thermally induced spin current in a S/F bilayer with a {\it homogeneous} exchange field without a DW because it is a main ingredient of the torque providing the spin pumping into the DW region. Let us apply a temperature difference $T_l-T_r$ to the ends of the film. In this case a thermally induced spin current appears  in the superconductor. This is a kind of a spin Seebeck effect. It is worth noting that the effect does not require an external spin source in the system as opposed to spin pumping experiments in superconductors \cite{Yao2018,Umeda2018}. The spin current in the homogenous S/F bilayer only carries $x$-spin component $J_x^x \equiv J$, which is directed along the ferromagnet magnetization. Fig.~\ref{bulkspin} demonstrates the dependence of the spin current on the system temperature  at small $\delta T = T_l - T_r \ll T$ for different $h$. For a homogeneous bilayer the spin thermopower  at $\delta T/\Delta_0 \ll 1$ is 
\begin{equation}
    \frac{2e^2 R_N J}{\delta T} = 
    F\left(\frac{\Delta+h}{2T}\right) - F\left(\frac{\Delta-h}{2T}\right)
\label{spin_current_hom_fin}
\end{equation}
with $F(x) = x \tanh x - \ln \cosh x$.
The maximal values of  $2eJR_N/\delta T$ are of the order of $(h/\Delta_0)\times 10^{-1}$mV/K and are reached for $T \sim 0.6-0.7 T_c$, as illustrated in Fig.~\ref{bulkspin}.

The estimated values of $2eJR_N/\delta T$ are much larger than that ones obtained for nonsuperconducting systems containing metallic ferromagnets. Such large values of the spin Seebeck effect are a result of the huge spin-dependent electron-hole asymmetry close to the Fermi level, see the inset of Fig.~\ref{bulkspin}. For the ballistic transport that we consider the distribution function of right-moving (left-moving) quasiparticles is determined by the Fermi distribution function of the left (right) end of the sample, which are assumed to be in thermal equilibrium at $T=T_{l(r)}$. Let us assume for simplicity that $T_r = 0$. Then there are no left-moving quasiparticles. The spin-split DOS occupied by right-moving quasiparticles is shown in the insert to Fig.~\ref{bulkspin}. We observe that at intermediate temperatures $\Delta-h<T<\Delta+h$ the spin-down DOS is presumably occupied by electron-like quasiparticles, while the spin-up DOS is occupied by hole-like quasiparticles. In this ideal situation all the thermally induced quasiparticles (both electrons and holes) have the same spin and contribute to the flow of spin-down quasiparticles to the right, which results in the maximal possible value of the thermally-induced spin current.

\begin{figure}[t]
    \centerline{\includegraphics[clip=true,width=3.5in]{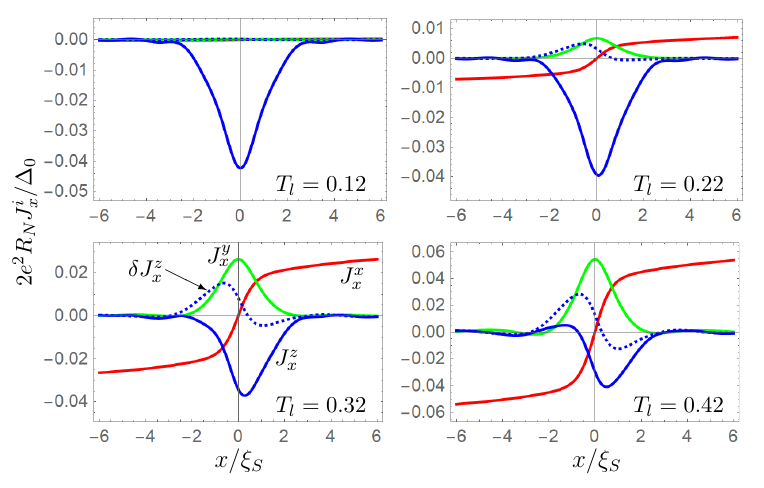}}
    \caption{Spatial profile of the spin current components $J_x^x$ (red), $J_x^y$ (green) and $J_x^z$ (blue) for different temperatures of the hot end, $\delta J_x^z=J_x^z-J_x^z(T_l=T_r)$.  $h_{eff}=0.3\Delta_0$, $T_r=0.02 \Delta_0$, $l_{DW}=0.5 \xi_S$, where $\xi_S = v_F/\Delta_0$ throughout the paper.}
 \label{thermospin_T}
 \end{figure}

{\it Thermally induced spin-transfer torque.}
The spatial profiles of the spin current in the presence of a plane DW (located in the $(x,y)$-plane) are presented in Fig.~\ref{thermospin_T} for different temperatures of the hot end. At first, let us focus on $J_x^x$ component, which is the only nonzero component of the thermally induced spin current in the bulk. Due to the presence of two magnetic domains with opposite magnetizations it leads to spin pumping into the region occupied by the DW. This process is schematically illustrated  on the top surface of Fig.~\ref{sketch} and is described there. At nonzero $T_l-T_r$ in-plane component $J_x^y$ also appears in the region of the DW. In the limit $T_l-T_r \to 0$ only $J_x^z$ survives. Then it represents  a spontaneous spin current occuring in the region occupied by the wall in equilibrium. It is carried by the equal-spin Cooper pairs generated by the magnetic texture. Similar spontaneous spin currents have already been obtained, usually in a Josephson junction-type geometry \cite{Bobkova2004,Grein2009,Alidoust2010,Shomali2011,Alidoust2015,Halterman2015,Jacobsen2016,konschelle2016,Aikebaier2018}.
The torque generated by this equilibrium spin current is compensated by the DW shape distortion resulting in additional contributions to the in-plane effective field (see supplemental material). Consequently, the equilibrium torque contribution does not effect the DW motion and is subtracted from nonequilibrium torque driving the wall. 

The torque can be obtained via the spin current according to Eq.~(\ref{torque}). In general, any spin torque can be written as $ \bm N = a~ \partial_x \bm m + b~ \bm m \times \partial_x \bm m $,
where $\bm m = \bm M/M$. The first (second) term can be related to electron spins following (being misaligned to) the magnetic texture. In the framework of the linear response theory the coefficients $a$ and $b$ are proportional to the temperature gradient.  For the plane DW under consideration $N_x$ and $N_y$ components contribute to the adiabatic torque and $N_z$ gives rise to the nonadiabatic contribution. 

The temperature dependence of the both adiabatic and nonadiabatic torques is determined by the spin pumping processes. It closely follows the temperature dependence of the bulk quasiparticle spin current (see supplemental material). Therefore, the spin pumping is the driving force of both adiabatic and nonadiabatic torque components. However, it is important to note that in the S/F hybrid the nonadiabatic torque naturally appears because of two different length scales: $l_{DW}$ and $\xi_S$. A quasiparticle  spin is aligned with the magnetization at the length scale $\sim \xi_S$. At $l_{DW} \lesssim \xi_S$ it inevitably mistracks magnetization giving rise to the nonadiabatic torque. The reason is that in the superconductor the quasiparticle is a coherent mixture of electron-like and hole-like excitations and strongly interacts with the condensate. Consequently, any changes of the quasiparticle spin are coupled to the changes in the equal-spin condensate wave function, which have characteristic spatial scale $\xi_S$.  This is in contrast to the nonsuperconducting case, where the nonadiabatic torque is believed to be due to spin-flip scattering processes.

{\it Thermally induced DW motion.}
The dynamics of the DW under the applied temperature difference is calculated from the LLG Eq.~(\ref{LLG}). At the present study we focus on small values of parameter $\zeta$ describing how strong is the torque induced by the superconductor. In this case we calculate the torque for the unperturbed DW neglecting the distortion of the DW shape due to its motion. Our numerical results for the spatial profiles of the moving DW are presented in the supplementary material and demonstrate that the distortion is indeed very small, therefore justifying the above assumption. 

\begin{figure}[t]
    \centerline{\includegraphics[clip=true,width=3.2in]{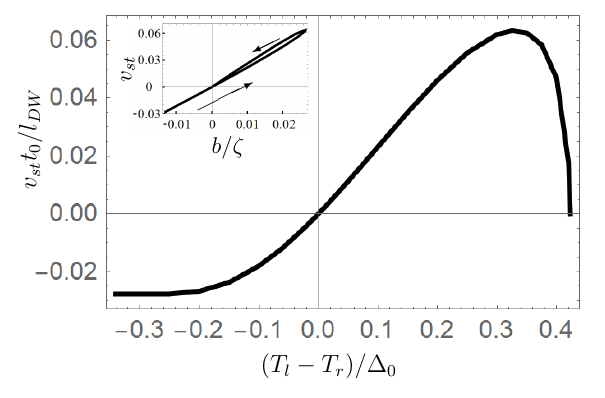}}
    \caption{DW velocity $v_{st}$ as a function of $\delta T = T_l-T_r$. $t_0 = (\gamma H_K)^{-1}$. Insert: $v_{st}$ as a function of $b(x=x_{DW})$. The direction of $\delta T$ growing along this curve is marked by the arrow. $\zeta=0.3$,  $\alpha = 0.2$, $K_\perp = H_K/M$, $T_r = 0.35 \Delta_0$.}
 \label{velocity}
 \end{figure} 

We found that for the values of $\zeta$ and $T_l-T_r$ considered in Fig.~\ref{velocity} the DW moves as a rigid object reaching the steady state at a characteristic time $t_d = 1/4 \pi \alpha \gamma M$, that is the Walker's breakdown \cite{Schryer1974} is not reached in our calculation. For the considered parameters we have found no sign of a precessional motion, which is typical for the motion in the regime after the Walker's breakdown. The DW velocity is calculated as $v(t) = \dot x_{DW}(t)$, where $x_{DW}(t)$ is the coordinate of the DW center at a given time and is extracted from the dynamical profiles of the magnetization. The steady state velocity $v_{st}=v(t \to \infty)$ as a function of $\delta T$ is plotted in Fig.~\ref{velocity}. We see that at $\delta T \ll \Delta_0$ the velocity is a linear function of the temperature difference. Although in superconducting systems the microscopically calculated  coefficients $a$ and $b$ are spatially dependent (see supplementary material for details), $v_{st} \sim b(x=x_{DW}) = - l_{DW}N_z(x=x_{DW})$, as it is demonstrated in the insert to Fig.~\ref{velocity}. It indicates that in this regime the DW motion is determined by the nonadiabatic torque analogously to the case of nonsuperconducting systems \cite{Li2006}.  
The "hysteretic behavior" of the parametric plot $v_{st}(b)$ is due to the nonmonotonic dependence of the velocity, as well as $b(x_{DW})$ on $\delta T$, which in turn results from the suppression of  superconductivity by heating of the film.  

The DW velocity $v_{st}$ is linearly proportional to the S/F coupling strength $\zeta$. At $\zeta=0.3$ and taking material parameters for Py films\cite{Beach2006} $H_K \sim 500$Oe and $l_{DW} \sim 20$nm  or for YIG thin films\cite{Mendil2019} $H_K \sim 0.5$Oe and $l_{DW} \sim 1\mu$m the maximal DW velocities can be estimated from Fig.~\ref{velocity} as $v_{Py} \sim 0.06 (\alpha/\alpha_{Py}) (l_{DW}/t_0)_{Py}$, what gives us  $v_{Py} \sim  100$m/s. Analogously, $v_{YIG} \sim 10^3$m/s. In these estimates we take into account that $v_{st} \sim \alpha^{-1}$ and realistic values of $\alpha_{Py} \sim 0.01$ and $\alpha_{YIG} \sim 10^{-4}$. 

{\it In summary,} we  have predicted  and microscopically calculated a thermally induced STT in thin film S/F bilayers containing a DW. It features adiabatic as well as nonadiabatic contributions. The physical mechanism of the torque is a unique feature of superconducting hybrids: it results from (i) the extremely efficient quasiparticle spin pumping into the superconducting region close to the DW provided by the giant Seebeck effect and (ii) strong interaction between quasiparticles and the condensate in the superconductor resulting in the characteristic length scale $\xi_S$ of the quasiparticle spin evolution, what, in its turn, gives rise to a nonadiabatic torque contribution at $l_{DW} \lesssim \xi_S$. We have demonstrated that this torque allows for a high-velocity steady DW motion corresponding to $v \sim 100$m/s at small temperature differences $\sim 1K$ applied at the length of several domain wall widths.

\begin{acknowledgments}
This work was financially supported by the Deutsche Forschungsgemeinschaft (DFG) through SFB 767 \textit{Controlled Nanosystems}. The work of A.M.B and I.V.B is carried out within the state task of ISSP RAS. I.V.B. also acknowledges the financial support by Foundation for the Advancement of Theoretical Physics and Mathematics “BASIS”.
\end{acknowledgments}

\section*{Supplemental Material}

\subsection*{Quasiclassical Keldysh Green's functions technique in terms of Riccati parametrization}

The matrix Green's function $\check g(\bm r, \bm p_F, \varepsilon, t)$ is a $8 \times 8$ matrix in the direct product of spin, particle-hole and Keldysh spaces and depends on the spatial vector $\bm r$, quasiparticle momentum direction $\bm p_F$, quasiparticle energy $\varepsilon$ and time $t$. In the S film it obeys the Eilenberger equation:
\begin{eqnarray}
 i \bm v_F \nabla\check g(\bm r, \bm p_F)+\Bigl[ \varepsilon \tau_z  + \bm h(\bm r) \bm \sigma \tau_z - \check \Delta,\check g \Bigr]_\otimes = 0,~~~~~~
 \label{eilenberger}
\end{eqnarray}
where $[A,B]_\otimes = A\otimes B -B \otimes A$ and $A \otimes B = \exp[(i/2)(\partial_{\varepsilon_1} \partial_{t_2} -\partial_{\varepsilon_2} \partial_{t_1} )]A(\varepsilon_1,t_1)B(\varepsilon_2,t_2)|_{\varepsilon_1=\varepsilon_2=\varepsilon;t_1=t_2=t}$. $\tau_{x,y,z}$ are Pauli matrices in particle-hole space with $\tau_\pm = (\tau_x \pm i \tau_y)/2$. $\hat \Delta = \Delta(x)\tau_+ - \Delta^*(x)\tau_-$ is the matrix structure of the superconducting order parameter $\Delta(x)$ in the particle-hole space.

In the ballistic case, it is convenient to use the so-called Riccati parametrization for the Green's function \cite{Eschrig2000,Eschrig2009}. In terms of the Riccati parametrization the retarded Green's function takes the form:
\begin{eqnarray}
\check g^{R,A} =
\pm N^{R,A} \otimes ~~~~~~~~~~~~~~~~~~~~~~~~~~~~ \nonumber \\ 
\left(
\begin{array}{cc}
1-\hat \gamma^{R,A} \otimes \hat {\tilde \gamma}^{R,A} & 2 \hat \gamma^{R,A} \\
2 \hat {\tilde \gamma}^{R,A} & -(1-\hat {\tilde \gamma}^{R,A} \otimes \hat \gamma^{R,A}) \\
\end{array}
\right),~~~~
\label{riccati_GF}
\end{eqnarray}

\begin{eqnarray}
\check g^{K} =
2N^R \otimes ~~~~~~~~~~~~~~~~~~~~~~~~~~~~~~~~~~~~~~~~~\nonumber \\
\left(
\begin{array}{cc}
x^K + \hat \gamma^{R} \otimes \hat {\tilde x}^K \otimes \hat {\tilde \gamma}^{A} & - (\hat \gamma^{R} \otimes \hat {\tilde x}^K - \hat x^K \hat \gamma^A) \\
\hat {\tilde \gamma}^{R} \otimes \hat x^K - \hat {\tilde x}^K \otimes \hat {\tilde \gamma}^{A}  & \hat {\tilde x}^K+\hat {\tilde \gamma}^{R} \otimes \hat x^K \otimes \hat \gamma^{A}) \\
\end{array}
\right) \otimes N^A ~~~~~~
\label{riccati_keldysh}
\end{eqnarray}
with
\begin{eqnarray}
N^{R,A} = \left(
\begin{array}{cc}
1+\hat \gamma^{R,A} \otimes \hat {\tilde \gamma}^{R,A} & 0 \\
0 & 1+\hat {\tilde \gamma}^{R,A} \otimes \hat \gamma^{R,A} \\
\end{array}
\right)^{-1}
\label{N}
\end{eqnarray}
where $\hat \gamma^{R,A}$, $\hat {\tilde \gamma}^{R,A}$, $\hat x^K$ and $\hat {\tilde x}^K$ are matrices in spin space. Note that our parametrization differs from the definition in the literature \cite{Eschrig2000,Eschrig2009} by factors $i\sigma_y$ as $\hat \gamma_{standard}^{R,A} = \hat \gamma^{R,A} i \sigma_y$ and $\hat {\tilde \gamma}_{standard}^{R,A} = i \sigma_y \hat {\tilde \gamma}^{R,A}$.
The Riccati parametrization Eqs.~(\ref{riccati_GF})-(\ref{N}) obeys the normalization condition $\check g \otimes \check g = 1$ automatically.

The Riccati amplitude $\hat \gamma$  obeys the following Riccati-type equations:
\begin{eqnarray}
 i \bm v_F \nabla \hat \gamma^R + 2 \varepsilon \hat \gamma^R = -\hat \gamma^R \otimes \Delta^* \otimes \hat \gamma^R - \bigl\{ \bm h \bm \sigma, \hat \gamma^R \bigr\}_\otimes - \Delta ~~~~~~
 \label{riccati}
\end{eqnarray}
and $\hat {\tilde \gamma}$ obeys the same equation with the substitution $\varepsilon \to -\varepsilon$, $\bm h \to -\bm h$ and $\Delta \to \Delta^*$.

The distribution function $\hat x^K$ obeys the equation:
\begin{eqnarray}
 i \bm v_F \nabla \hat x^K + i \partial_t \hat x^K + \hat \gamma^R \otimes \Delta^* \otimes \hat x^K + \nonumber \\
 \hat x^K \otimes \Delta \otimes \hat {\tilde \gamma}^A + [\bm h \bm \sigma,\hat x^K]_\otimes = 0 ,
 \label{distribution}
\end{eqnarray}
while $\hat {\tilde x}^K$ obeys the same equation with the substitution $\bm h \to -\bm h$, $\Delta \to \Delta^*$, $\hat \gamma^{R,A} \leftrightarrow \hat {\tilde \gamma}^{R,A}$. In this work, we assume $\Delta = \Delta^*$.

If we consider a locally spatially inhomogeneous  magnetic texture like a domain wall, the Riccati amplitudes $\hat \gamma$ and $\hat {\tilde \gamma}$ can be found from Eq.~(\ref{riccati}) numerically with the following asymptotic condition:
\begin{align}
 \hat \gamma_{\infty}  = & \gamma_{0\infty} + \frac{\bm h_{\infty} \bm \sigma}{h} \gamma_\infty , \\
 \gamma_{0\infty}  = & -\frac{1}{2}\Bigl[ \frac{\Delta}{\varepsilon +h+i\sqrt{\Delta^2 - (\varepsilon + h)^2}} \nonumber \\ 
 & + \frac{\Delta}{\varepsilon-h+i\sqrt{\Delta^2 - (\varepsilon  - h)^2}} \Bigr],
 \\
 \gamma_\infty = & -\frac{1}{2}\Bigl[ \frac{\Delta}{\varepsilon+h+i\sqrt{\Delta^2 - (\varepsilon  + h)^2}} \nonumber \\ & - \frac{\Delta}{\varepsilon -h+i\sqrt{\Delta^2 - (\varepsilon  - h)^2}} \Bigr],
  \label{riccati_asymptotic}
\end{align}
and $\hat {\tilde \gamma}_\infty = - \hat \gamma_\infty$.

Eq.~(\ref{riccati}) is numerically stable if it is solved starting from $x = -\infty$ for right-going trajectories $v_x > 0$ and from $x = +\infty$ for left-going trajectories $v_x < 0$. On the contrary, $\hat {\tilde \gamma}$ can be found numerically starting from $x = +\infty$ for right-going trajectories $v_x > 0$ and from $x = -\infty$ for left-going trajectories $v_x < 0$. The advanced Riccati amplitudes can be found taking into account the relation \cite{Eschrig2009} $\hat \gamma^A = -(\hat {\tilde \gamma}^R)^\dagger$. The superconducting order parameter is to be found self-consistently according to 
\begin{eqnarray}
\Delta = -\frac{\lambda}{8} \int \limits_{-\Omega}^\Omega d \varepsilon {\rm Tr}_4\langle \tau_- \check g^K \rangle,
\label{self_con}
\end{eqnarray}
where $\langle ... \rangle$ means averaging over the Fermi surface, and ${\rm Tr}_4$ is the trace in Nambu$\otimes$Spin space. $\lambda$ is the coupling constant and $\Omega$ is the Debye frequency cutoff.

If we neglect the dependence of $\bm h$ on time, then it follows from Eq.~(\ref{distribution}) that the distribution function $\hat x^K$ for a given ballistic trajectory is determined by the equilibrium distribution function  of the left (right) reservoir for $v_{F,x} >0$ ($v_{F,x} <0$) and takes the form
\begin{eqnarray}
\hat x^K_{\pm} = (1+\hat \gamma^R_\pm \otimes \hat {\tilde \gamma}^A_\pm)\tanh \frac{\varepsilon}{2T_{l,r}},
\label{distrib_eq}
\end{eqnarray}
where the subscript $+(-)$ corresponds to the trajectories $v_{F,x}>0$ ($v_{F,x}<0$). On the contrary,
\begin{eqnarray}
\hat {\tilde x}^K_{\pm} = -(1+\hat {\tilde \gamma}^R_\pm \otimes \hat \gamma^A_\pm)\tanh \frac{\varepsilon}{2T_{r,l}}.
\label{tilde_distrib_eq}
\end{eqnarray}
The terms $\propto \dot {\bm h}$ in Eq.~(\ref{distribution}) can be neglected under the conditions $(h/\Delta)v_{st}/l_{DW}\Delta \ll 1$ and $(h/\Delta)(1/t_d \Delta) \ll 1$, where $t_d$ is the characteristic time of the induced magnetization dynamics, $v_{st}$ is the characteristic DW velocity and $l_{DW}$ is the DW width. For realistic parameters $t_d \sim 10^{-9}-10^{-8}c$, $v_{st} \sim 100 m/s$ according to our estimates. Therefore, at $\Delta \sim 1K$ and $h/\Delta \lesssim 1$ these conditions are fulfilled to a good accuracy for any experimentally reasonable DW width $l_{DW} \sim 10nm - 1\mu m$. Physically, these terms  account for the electromotive force, which arises in the system due to the magnetization dynamics and has been studied in different contexts before \cite{Stern1992,Stone1996,Volovik1987,Berger1986,Barnes2007,Duine2008,Saslow2007,Tserkovnyak2009,Zhang2009,Yang2009,Yang2010,Rabinovich2019,Rabinovich2019_2}, but here its back influence on the magnetization dynamics can be safely neglected. 

The quantity of the main interest for us is the spin current flowing in the superconductor. It exerts a torque on the ferromagnet magnetization. The spin current of spin projection $\bm J$ in direction $j$ can be calculated as follows:
\begin{eqnarray}
\bm J_j = -\frac{ N_F}{16} \int \limits_{-\infty}^\infty d \varepsilon {\rm Tr}_4 \Bigl[ \bm \sigma \langle v_{F,j} \check g^K \rangle \Bigr],
\label{spin_current}
\end{eqnarray}
where $\check g^K(\epsilon,\bm v_F)$ represents the Keldysh part of the quasiclassical Green's function. $N_F$ is the normal density of states at the Fermi level.

\subsection*{Spontaneous spin current, DW magnetization profile and self-consistent superconducting order parameter in equilibrium S/F bilayer}

 \begin{figure}[t]
    \centerline{\includegraphics[clip=true,width=3.2in]{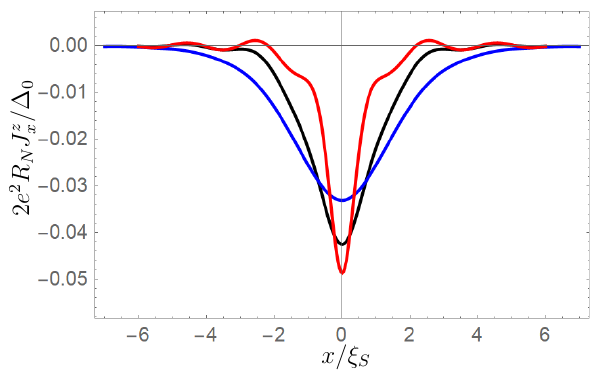}}
    \caption{Spin current $J_x^z(x)$ at the plane DW in equilibrium. Other components of the spin current are zero.  $l_{DW}=\xi_S$(blue); $0.5 \xi_S$(black); $0.2\xi_S$(red).  $h_{eff}=0.3\Delta_0$, $T=0.02\Delta_0$.}
 \label{spcur}
 \end{figure}

 \begin{figure}[t]
    \centerline{\includegraphics[clip=true,width=3.5in]{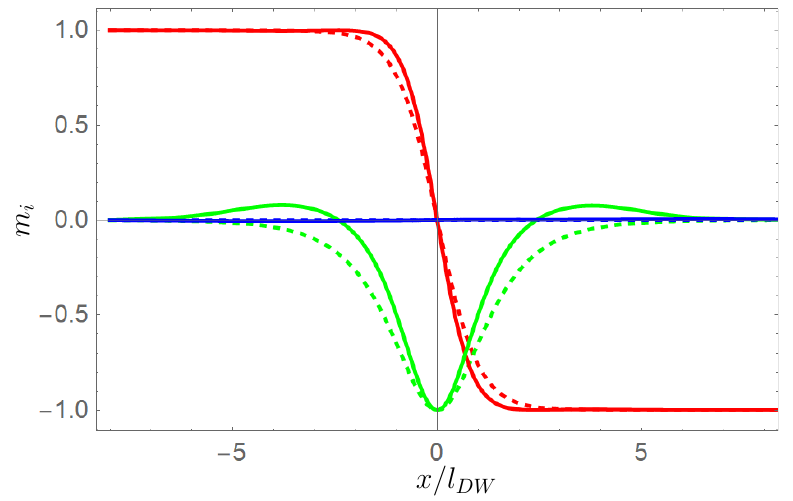}}
    \caption{Magnetization profile in the equilibrium S/F bilayer for $\zeta=10$. $\bm m_i = \bm M_i/M$. Different magnetization components are plotted in different colors: $m_x$-red, $m_y$-green and $m_z$-blue. The dotted curves correspond to $\zeta=0$ (the superconductor is absent). $l_{DW}=0.5\xi_S$, $h_{eff}=0.3\Delta_0$, $T=0.02 \Delta_0$, $K_\perp = H_K/M$.}
 \label{equilibrium_profile}
 \end{figure}

The spontaneous spin current for our ballistic S film in proximity to the ferromagnet with a coplanar DW is plotted in Fig.~\ref{spcur}.
 It is seen that the amplitude of the spontaneous current is higher for narrow DWs. In the limit $l_{DW}/\xi_S \gg 1$ it disappears. 
The spin current is not conserved and exerts a spin-transfer torque on the magnetization. We find the resulting equilibrium shape of the DW accounting for the spin-transfer torque from the LLG equation. It is found that the presence of the superconductor results in the appearance of the additional oscillations of the magnetization in the $(x,y)$-plane. These oscillations generate additional contributions to the in-plane effective field, which exerts a $z$-directed torque on the magnetization, thus compensating the action of the spontaneous spin current. The resulting magnetization profile in equilibrium is presented in Fig.~\ref{equilibrium_profile}. For realistic ratios of the anisotropy field to the demagnetization field $H_K/K_\perp M$ we have found no noticeable deviation of the DW shape from the initial $(x,y)$-plane. At the same time the distortion of the DW is accompanied by its narrowing, which also appears to provide appropriate contributions to the in-plane effective field. It is worth noting that Fig.~\ref{equilibrium_profile} is plotted for the extremely high value of $\zeta=10$ to make the distortions clearly visible.

Equilibrium S/F bilayer with a DW was previously considered for a dirty system in Ref.~\onlinecite{Aikebaier2018} based on the free energy consideration. For a trial Neel-type plane DW it was found that the presence of the superconductor shrinks the DW size. This effect is closely connected to the fact that the the superconductivity can be enhanced for narrow DWs due to the effective averaging of the exchange field, and, consequently, to weaker suppression of superconductivity in the DW region. The increase of the superconducting order parameter provides the corresponding gain in the condensation energy. Here we do not see this effect because the considered exchange fields and temperatures are too small to cause essential suppression of the order parameter far from the DW, what is a necessary condition to have the superconductivity enhancement (restoring) near the DW. Instead, we observe weak Friedel-like oscillations of the order parameter near the DW and suggest that it is a specific feature of the ballistic limit we consider.

  \begin{figure}[t]
    \centerline{\includegraphics[clip=true,width=3.2in]{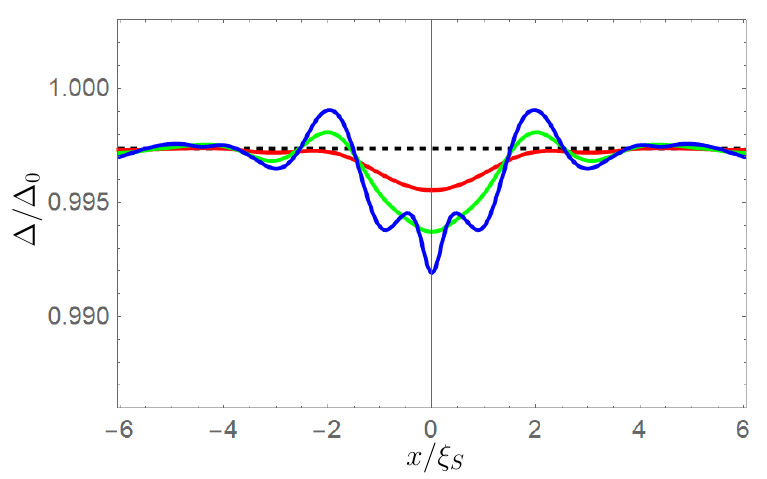}}
    \caption{Self-consistent order parameter as a function of the spatial coordinate along the bilayer. $l_{DW}=\xi_S$(red); $0.5 \xi_S$(green); $0.2\xi_S$(blue). $h_{eff}=0.3\Delta_0$. Black dotted line represents the order parameter value in the S/F bilayer in the absence of a DW.}
 \label{selfconsist}
 \end{figure}
 
The self-consistent profile of the order parameter calculated in the presence of a DW in the F layer is demonstrated in Fig.~\ref{selfconsist} for three different DW widths. It is seen that it manifests Friedel-like oscillating behavior. The oscillations become more pronounced for narrow DWs, but are generally weak for considered values of the suppression factors: exchange field and temperature.

\subsection*{Details of the order parameter and torque calculations under the applied temperature difference}

\begin{figure}[t]
    \centerline{\includegraphics[clip=true,width=3.2in]{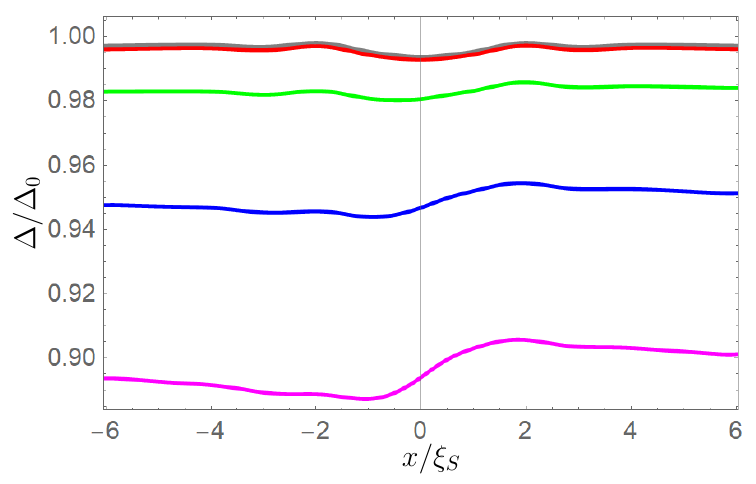}}
    \caption{Self-consistent profile of the order parameter in the presence of the DW and a temperature difference. Different curves correspond to different temperatures of the hot end $T_l = 0.02 \Delta_0$(grey), $0.12$(red), $0.22$(green), $0.32$(blue) and $0.42$(purple). $h_{eff}=0.3~\Delta_0$, $T_r=0.02\Delta_0$, $l_{DW}=0.5 \xi_S$.}
 \label{self_delta}
 \end{figure} 

At first in Fig.~\ref{self_delta} we demonstrate   results for the self-consistent profile of the superconducting order parameter in the presence of the DW and a temperature difference. The small overall suppression of the order parameter by heating of the superconductor is clearly seen. More interesting feature is that the order parameter is additionally suppressed  near the DW from the "hotter" side and slightly enhanced with respect to the bulk value from the "colder" side of the DW. It seems that there appears an excess (lack) of  quasiparticles at the corresponding side of the DW.

\begin{figure}[t]
    \centerline{\includegraphics[clip=true,width=3.5in]{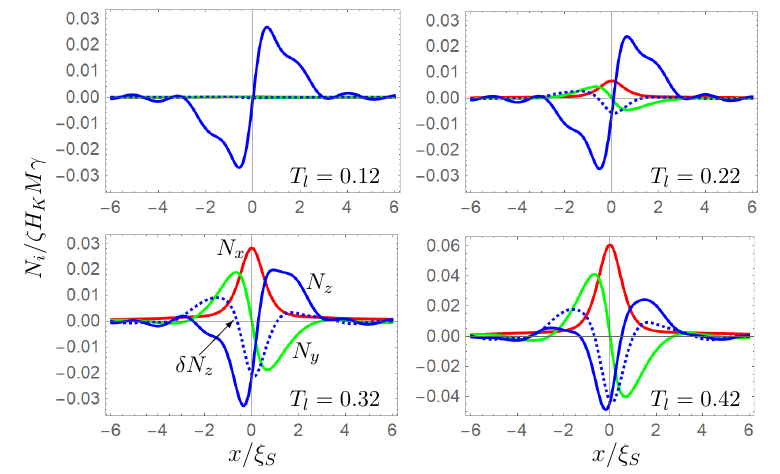}}
    \caption{Spatial profile of the torque components $N_x$ (red), $N_y$ (green) and $N_z$ (blue) at the plane DW under the applied heat bias, $\delta N_z = N_z - N_z(T_l=T_r)$. $h_{eff}=0.3\Delta_0$, $T_r=0.02 \Delta_0$, $l_{DW}=0.5 \xi_S$.}
 \label{torque_T}
 \end{figure}
 
\begin{figure}[!tbh]
 \begin{minipage}[b]{\linewidth}
   \centerline{\includegraphics[clip=true,width=2.8in]{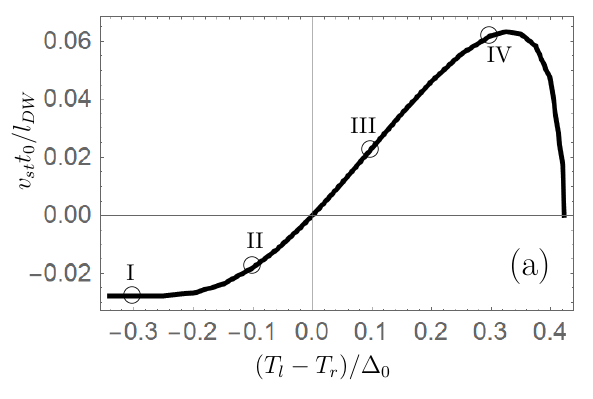}}
   \end{minipage}
   \hfill
   \begin{minipage}[b]{\linewidth}
   \centerline{\includegraphics[clip=true,width=2.8in]{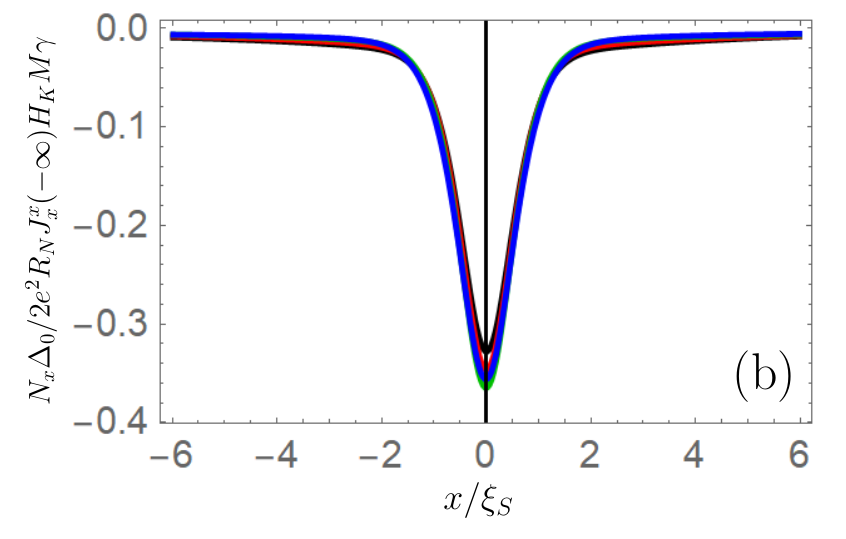}}
   \end{minipage}
   \hfill
   \begin{minipage}[b]{\linewidth}
   \centerline{\includegraphics[clip=true,width=2.8in]{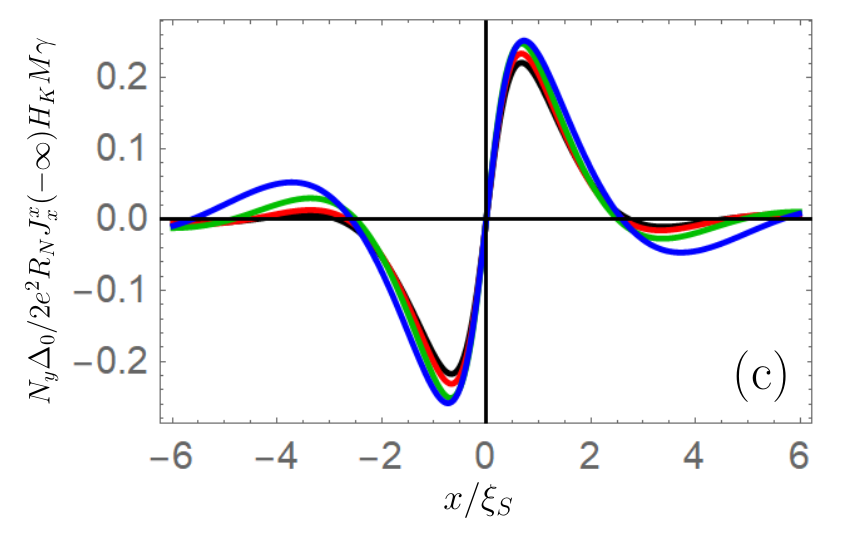}}
   \end{minipage}
   \hfill
   \begin{minipage}[b]{\linewidth}
   \centerline{\includegraphics[clip=true,width=2.8in]{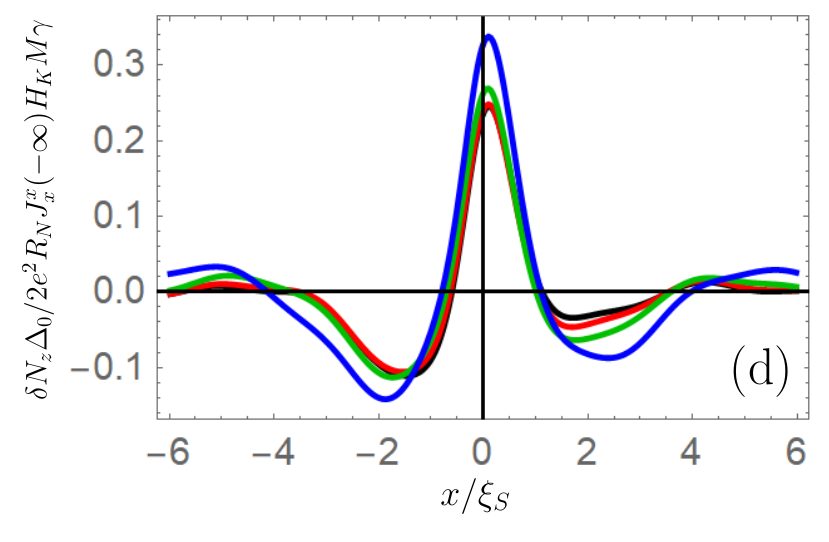}}
   \end{minipage}
      \caption{(a) DW velocity as a function of the temperature difference. All the parameters are the same as in the main text. (b)-(d) Spatial profiles of the torque components, normalized to the bulk thermally-induced spin current. Different curves correspond to different temperature differences, marked by circles in panel (a): I-black; II-red; III-green; IV-blue. }
 \label{torque_velocity}
 \end{figure}

The spatial profiles of the torque $\bm N$ acting on the DW from the spatially-dependent spin current, shown in Fig.~3 of the main text, are presented in Fig.~\ref{torque_T}. The torque exhibit strong temperature dependence. In order to investigate this dependence in more detail, in Figs.~\ref{torque_velocity}(b)-(d)  we present the torque components at different temperatures normalized to the value of the bulk thermally-induced spin current at a given temperature difference. It is seen that the temperature dependence of the torque closely follows the temperature dependence of the bulk thermally-induced spin current, what means that the torque is indeed driven by the thermal spin pumping process, described in the caption to Fig.1 of the main text. Panels(b)-(d) of Fig.~\ref{torque_velocity} represent all three torque components. The particular temperature differences are encoded by colors and marked by the circles in Fig.~\ref{torque_velocity}(a). 

In order to investigate the spatial structure of the torque induced by the presence of the superconductor, we separate it to adiabatic $ a \partial_x \bm m $ and nonadiabatic $b\bm m \times \partial_x \bm m $ contributions. In fact, the full microscopic result obtained from Eq.~(5) of the main text also contains the term $\sim \bm m$, but we exclude this contribution because in the framework of the considered model the amplitude of the magnetization is fixed.

\begin{figure}[t]
    \centerline{\includegraphics[clip=true,width=3.5in]{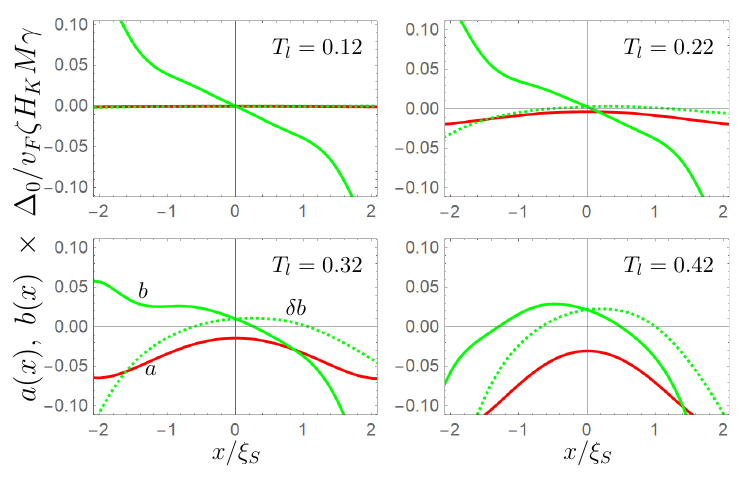}}
    \caption{Torque coefficients $a(x)$(red) and $b(x)$(green) for different temperatures of the hot end, $\delta b = b - b(T_l=T_r)$. $h_{eff}=0.3\Delta_0$, $T_r=0.02 \Delta_0$, $l_\textrm{DW}=0.5 \xi_S$.}
 \label{coefficients}
 \end{figure}
 
\begin{figure}[t]
    \centerline{\includegraphics[clip=true,width=3.0in]{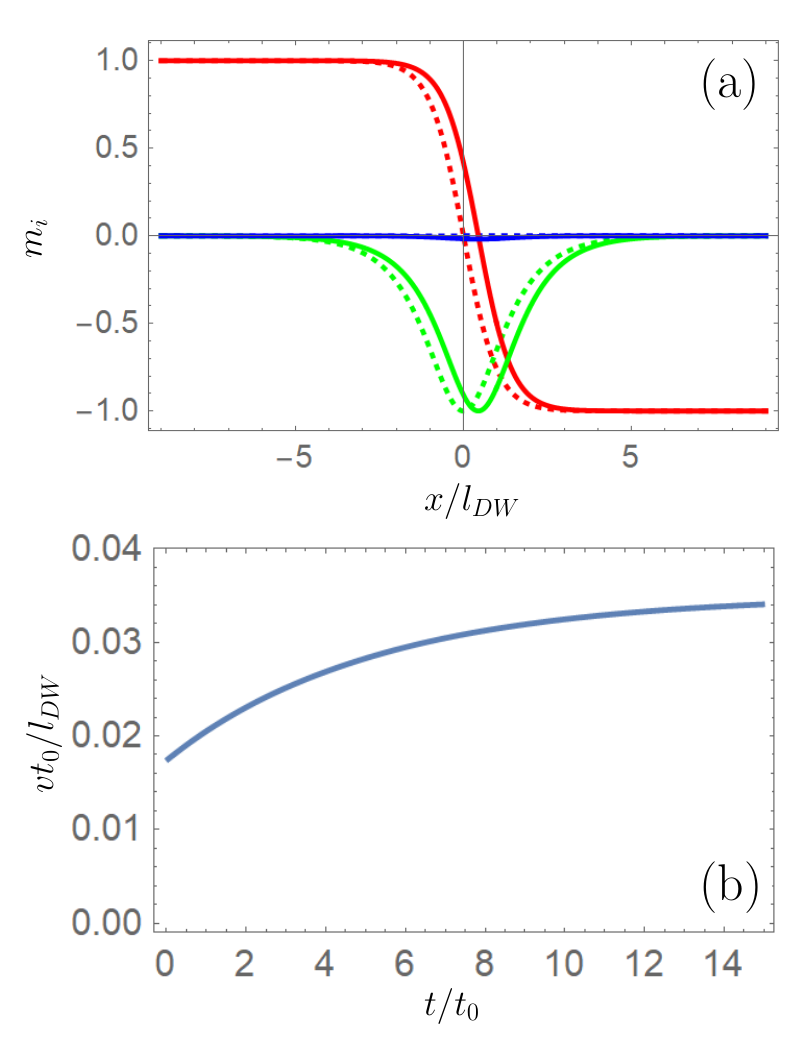}}
    \caption{(a) Components of the magnetization profile at $t=0$ (dashed lines) and at $t=15t_0$ (solid lines). $m_x$-red; $m_y$-green and $m_z$-blue. (b) the DW velocity as a function of time. $\zeta=0.3$,  $\alpha = 0.2$, $K_\perp = H_K/M$, $T_r = 0.35 \Delta_0$ and $T_l=0.50 \Delta_0$.}
 \label{dynamics}
 \end{figure} 

The essential feature of the superconducting system is that the microscopically calculated  coefficients $a$ and $b$ in our case are spatially dependent. They are plotted in Fig.~\ref{coefficients} as functions of $x$-coordinate (the DW center is at $x=0$) for different temperatures of the hot end. It is different from the nonsuperconducting case, where they are typically do not depend on coordinates due to absence of a corresponding spatial scale.  Here the characteristic scale of the spatial variation of $a$ and $b$ is determined by the superconducting coherence length $\xi_S$. It is interesting that the coefficient $b$ is even a sign-changing function of the $x$-coordinate. In equilibrium, at $T_l = T_r$, $b = 0$ at $x = x_{DW}$, that is, in the center of the DW. Under the applied temperature difference it is useful to introduce $\delta b = b - b(T_l = T_r)$. It becomes nonzero at $x=x_{DW}$: $\delta b(x=x_{DW}) = b(x=x_{DW}) \neq 0$. The DW velocity is proportional to this quantity as it is demonstrated in the main text.

\subsection*{Magnetization dynamics and details of the DW velocity calculations}

Fig.~\ref{dynamics} represents all the three Cartesian components of the magnetization profile as functions of $x$-coordinate at the initial moment $t=0$ when the tempetature difference is applied and at some $t_f >0$ when the DW motion practically reaches stationary regime. After examining the dynamics between $t=0$ and $t_f$ we concluded that the DW moves as a rigid object keeping its initial shape to a good accuracy. Therefore, we are in the regime before the Walker's breakdown. Fig.~\ref{dynamics}(a) demonstrates that the distortions of the DW shape are indeed negligible. 

In the regime below Walker's breakdown the DW center position $x_{DW}$ is a well-defined quantity and the DW velocity can be extracted from the numerically calculated magnetization profiles as  $v = \dot x_{DW}$. The DW velocity $v$ as a function of time is presented in Fig.~\ref{dynamics}(b).

\end{document}